\documentclass[graybox]{svmult}

\usepackage{mathptmx}       
\usepackage{helvet}         
\usepackage{courier}        
\usepackage{type1cm}        
\usepackage{makeidx}         
\usepackage{graphicx}        
\usepackage{multicol}        
\usepackage[bottom]{footmisc}

\makeindex             

\begin{document}

\title*{Gould's Belt}
\author{Jan Palou\v s 
and So\v na Ehlerov\' a
}
\institute{Jan Palou\v s \at Astronomical Institute, Academy of Sciences of the Czech Republic, Bo\v cn\' \i \ II, 141 00 Prague 4, Czech Republic \email{palous@ig.cas.cz}
}
%
%
\maketitle

\abstract{The local velocity patterns of star forming regions, young OB stars, nearby OB associations, atomic and molecular gas are confronted with models of an expanding region. We test free expansion from a point or from a ring, expanding 2D shell, and expanding 3D belt with abrupt or gradual energy injection snow-plowing the ambient medium with or without the drag forces including fragmentation and porosity of the medium. There is no agreement on the expansion time, which varies from 30 - 100 Myr. The inclination of the Gould belt is not explained by the above models of expansion. An oblique impact of a high velocity cloud may explain it, but the observed velocity pattern is difficult to reproduce. The Gould's belt may be one of the many structures resulting from shell-shell collisions in the galactic plane. The origin of the Gould's belt may be connected to instabilities in the curling gas flows downstream from the Galaxy spiral arms, forming ISM clouds and  star formation complexes.}


\section{Introduction}
\label{sec:1}


Gould's belt is a flat system of young OB stars inclined $\sim 20^{o}$ to the  plane of the Milky Way, with the line of nodes pointing approximately in the direction  of galactic rotation. In the direction to the galactic center, it extends up to $\sim 250$ pc from the Sun inclining towards positive $z$. In the opposite direction to the Galaxy anticenter, it reaches up to  $\sim 500$ pc from the Sun inclining towards negative $z$. Nearby Scorpius-Centaurus, Perseus and Orion OB associations together with other associations in Vela, Scutum and Vulpecula form an elliptical belt-like structure with center in the second galactic quadrant at a distance $\sim$ 100 pc from the Sun. Motions of young stars in the Gould's belt deviate from galactic differential rotation. It is a challenge to explain their space distribution and kinematics using a plausible model. 

Besides stars, also HI gas feature A was discovered by Lindblad (1967). Later, it was designated as Lindblad's ring, which has a similar space distribution as OB stars. Its kinematics was analysed by Lindblad et al. (1973) and explained as an expanding HI shell. Also nearby CO and H$_2$ clouds showing a local hole may be parts of the Gould's belt (Dame et al. 1987; Taylor et al. 1987; see also Perrot \& Grenier, 2003). A comprehensive overview of the Gould's belt is given by  P\" oppel(1997): space distribution and kinematics of local young OB stars, HI, H$_2$ dark clouds and star forming regions are interpreted with two basic models: as a result of an explosive event connected to winds and SN in a young OB association, or as a consequence of a collision of a high velocity cloud (HVC) with the galactic plane.   

Here we discuss the kinematics of local young OB stars, HI and local star forming regions, and we try to include it into the large-scale picture of galaxy flows compressing the ISM clouds and triggering star formation down-stream after passage through a galaxy spiral arm.    


%

\section{OB stars}

\begin{center}
\begin{figure}[t]
\includegraphics[scale=.45,angle=0]{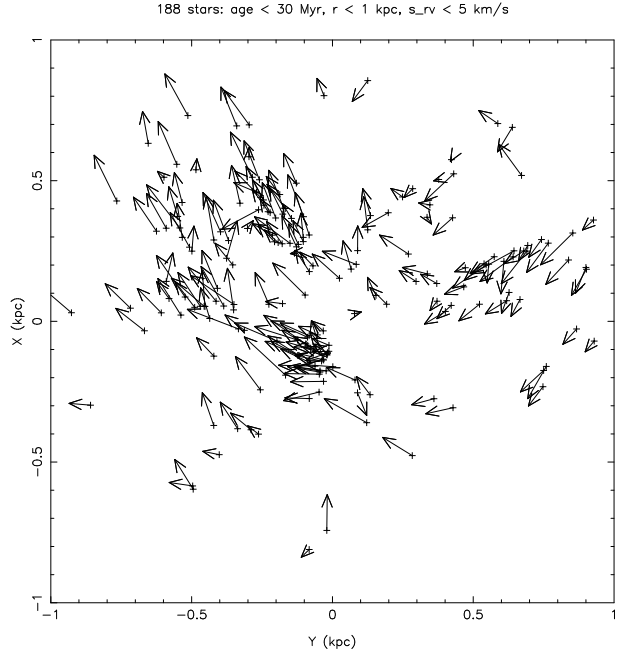}
%
%
\caption{Positions and space velocity vectors of young OB stars within 1 kpc from the Sun.}
\label{fig1}       
\end{figure}
\end{center} 

\noindent
Radial velocities $V_r$, Hipparcos proper motions $\mu_l, \mu_b$ and distances $r$ derived using the Str\" omgren photometry may be used in condition equations describing the linearized approximation of the local velocity field:
\begin{eqnarray}
V_r =  U_{\odot } \cos l \cos b - V_{\odot } \sin l \cos b - W_{\odot } \sin b \cr
 + r \cos^2 b~(K + A \sin 2l + C \cos 2l)   \cr
 + r~(H \sin^2 b + D \cos l \sin 2b + G \sin l \sin 2b) 
\label{vr}
\end{eqnarray}
\begin{eqnarray}
4.74~r~\mu_l   \cos   b  =  - U_{\odot }  \sin  l -
V_{\odot }  \cos  l \cr
 + r  \cos  b ~( B + A  \cos  2l - C  \sin  2l )  \cr
 + r  \sin  b  \cos  b  ~\lbrack (E - D) \sin  l \cr 
 + (G -  F) \cos  l \rbrack 
\label{mul}
\end{eqnarray}
\begin{eqnarray}
4.74 ~r ~\mu_b =  - U_{\odot }  \cos  l  \sin  b +
V_{\odot }   \sin  l  \sin  b -
W_{\odot}  \cos  b \cr
  - r  \sin  b  \cos  b ~\lbrack ( K - H ) + A  \sin  2l
+ C  \cos  2l  \rbrack \cr
+ 2r  \sin^2 b ~\lbrack ( E - D )   \cos  l
+ ( F - G ) \sin  l \rbrack 
\label{mub}
\end{eqnarray}
$l, b$ are the galactic longitude and latitude and
\begin{eqnarray}
A &= &-{1 \over 2}\left( {\partial U \over \partial Y} + {\partial V \over \partial X}\right) \\
B &= &{1 \over 2}\left( {\partial U \over \partial Y} - {\partial V \over \partial X}\right) \\
C &= &{1 \over 2}\left( {\partial U \over \partial X} - {\partial V \over \partial Y}\right) \\
K &= &{1 \over 2}\left( {\partial U \over \partial X} + {\partial V \over \partial Y}\right) \\
H &= &{\partial W \over \partial X} \\
D &= &{1 \over 2}{\partial W \over \partial Y}  \\
G &= &{1 \over 2}{\partial W \over \partial Z}  \\
E &= &{1 \over 2}\left( {\partial U \over \partial Z} -
{\partial W \over \partial X} \right) \\
F &= &-{1 \over 2}\left( {\partial V \over \partial Z} -
{\partial W \over \partial Y} \right).  
\end{eqnarray}
are the parameters describing the local velocity field. X, Y, and Z are the coordinates centered on the Sun pointing away from the galactic center, in the direction of the galactic rotation and perpendicularly to the galactic plane, and U, V, W are the components of the space velocity vectors along X, Y, and Z. 

(U, V) components of the space velocity vectors of the OB stars within 1 kpc from the Sun and younger than 30 Myr   
are plotted in Fig. \ref{fig1}. We see that the differential galactic rotation of young stars around the distant center  is locally perturbed by wave-like motion following more complex galactic orbits.  
Outside the Gould's belt, the condition equations (\ref{vr}), (\ref{mul}) and (\ref{mub}) for young OB stars give $A = 12.8$ km s$^{-1}$kpc$^{-1}$, $B = -12.5$ km s$^{-1}$kpc$^{-1}$, and C, K, H, D, G, E and F close to zero, which may be interpreted as a consequence of rotation following a flat rotation curve with a circular angular velocity at the Sun of $\sim $25 km s$^{-1}$ kpc$^{-1}$, and no significant deviations from circularity when the Gould's belt region is excluded. 

However, the solution of equations (\ref{vr}), (\ref{mul}), and (\ref{mub})
for Gould's belt stars identified due to their position in the local flat inclined system, as it was proposed by  
Westin (1985), Palou\v s (1985), Comeron, (1994) and Lindblad et al., (1997), presents significant kinematical deviation from galactic differential rotation. It shows that the Gould's belt moves away from the galactic center at about $\sim 5$ km s$^{-1}$. Large negative $B$ and  large positive $K$ result from  additional rotation around a local center combined with expansion from it. The values of $H$ and $D$ are also significantly different from 0 showing a gradient in $W$ motions of $\sim 7$ km s$^{-1}$kpc$^{-1}$. The average $W$ motion is proportional to the distance from the rotation axis pointing in the direction $\sim 340^o$ (Comeron, 1999). This shows that $W$ motions of the Gould's belt stars are coherent, however, the kinematic axis of $W$ motions is displaced from the line of nodes between Gould's belt and the galactic plane.   

\section{Expansions}

Expansion from a small region was examined since Blaauw (1952), who described the evolution of an expanding group in the galactic plane. Later, the model of an expanding group was applied to local O and B type stars by Lesh (1968, 1972a, 1972b), who derived the plane parallel velocity gradients. Lindblad et al. (1973)  used the free expansion model to reproduce the HI ring around the Sun: they concluded that the initial expansion velocity was 3.6 km/s and the expansion age was close to 60 Myr. The velocity field of free expansion from a small region (Lindblad 1980) shows a positive  $K$ similar to the observed value. However, $B$ is time independent and equal to 0, which is far from the large negative value derived from the velocity field of Gould's belt stars. Later, the free expansion model was modified with initial rotation (Palou\v s 1998a, 1998b), and expansion was decelerated in a model of an expanding supershell by Olano (1982). A model with 3D snow-plough in the differentially rotating galaxy was proposed by Ehlerov\' a et al. (1997),  Moreno (1999) used the snow-plough in 2D and compared its plane parallel shape to the distribution of OB associations in the Gould's belt. A supershell forming dense belt expanding in 3D was examined by Perrot \& Grenier (2003), who tested gradual energy injection, pure snow-plough, pure drag without accretion, fragmentation of the shell with increasing porosity and triggering star formation in pre-existing clouds. However, a common property of  all these models of expansion is the small value of the constant $B$, which does not correspond to observations. 
The inclination of the Gould's belt and its size, which is much larger compared to the thickness of the galactic disk, remains also unexplained. In models by Perrot \& Grenier (2003) an expanding inclined belt forms out of a gaseous disk, which has a pre-existing tilt relative to the galactic disk.     

\begin{center}
\begin{figure}[h]
\includegraphics[scale=.45,angle=0]{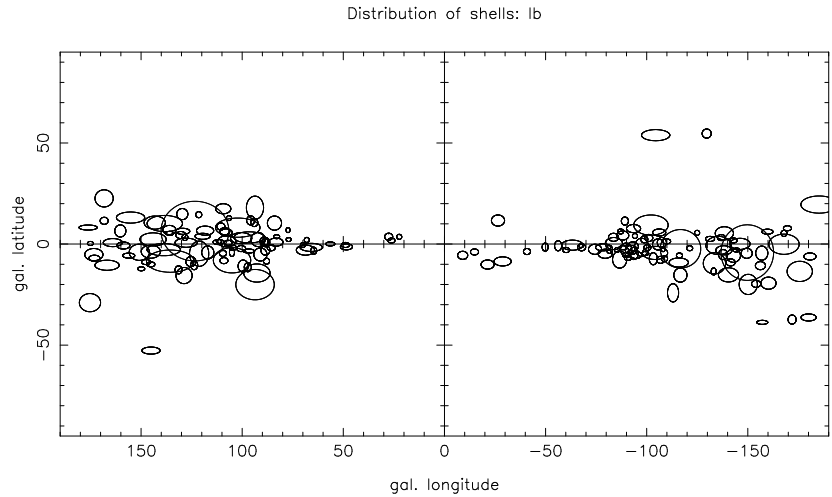}
%
%
\caption{HI shells in the Milky Way: the inner Galaxy and local shells are excluded.}
\label{fig2}       
\end{figure}
\end{center}

\section{Impacts}

\noindent
Some of the shells and supershells observed in HI and CO in the Milky Way and nearby galaxies including M31 may have been formed by an impact of a high velocity cloud (HVC) into the Galaxy ISM. Tenorio-Tagle (1980) argues that the kinetic energy of a HVC corresponds to the energy requirements for formation of a supershell in a galactic disk. They show (Tenorio-Tagle et al., 1986) that the sizes of structures created by an HVC impact have radii of a few hundreds of parsecs and expanding velocities of 10 km/s or more. Gould's belt may have been formed by such HVC versus galactic disk collision, similar to some of the HI holes observed in nearby galaxies.
 
An impact of a HVC with a velocity component parallel to the galactic plane was studied by Comeron \& Torra (1992). They show that such collision creates a shock-wave with some inclination to the Galaxy symmetry plane. An oblique impact of a HVC may explain the Gould's belt. Later, Comeron \& Torra (1994) used an oblique impact to derive the observed  gradients of the $W$ velocity component. However, a full 3D hydrodynamic treatment of the supersonic impact of a HVC into the ISM of a galaxy including radiative transfer and all cooling and heating processes is still to be done.  

\section{Galactic orbits}

With positions and space velocity vectors, we may compute the galactic orbits of nearby young stars and explore how the shape of the current local system changes going to the past. If the Gould's belt has a common origin, its imprint should be recorded in orbits. Computing the orbits backwards in time may discover the shape of the original region where it was born.  

The volume taken by members of Scorpius-Centaurus associaton gets smaller when we go to the past (Palou\v s 1998a, b). The smallest volume occupied by members of present OB associations in Lower Centaurus-Crux, Upper Centaurus-Lupus and Upper Scorpius is reached 10 - 12 Myr ago. At that time, the progenitors of current OB stars including Scorpius-Centaurus and Orion associations form a sheet-like region about 500 pc long and less than 100 pc wide with the main axis pointing in the direction l = 20 - 200$^o$. 

The local ISM was explored by Frisch (1981), Frich and York (1983) and Heiles (1989), who describe the Local Bubble, which interacts with Loop I Bubble and Loop IV  Bubble above the plane. A supersonic collision of two expanding supershells may create a sheet-like region composed of matter collected in both colliding shells. Formation of molecular clouds and subsequent star formation may have been triggered due to the shell-shell collision. Later, the young OB stars have expanded to the present configuration of the Gould's belt. The ISM sheet could be tilted relative to the galactic plane, which is a natural outcome of collisions between off plane asymmetrical supershells. 3D hydrodynamical simulations of shell - shell supersonic collisions, mass accumulation in a sheet-like region, formation of dense molecular clouds, and triggering of star formation is to be performed in future to explore this possibility that the Gould's belt may have been formed this way.

\section{Are there other Gould's Belts}

\begin{center}
\begin{figure}[h]
\includegraphics[scale=.45,angle=270]{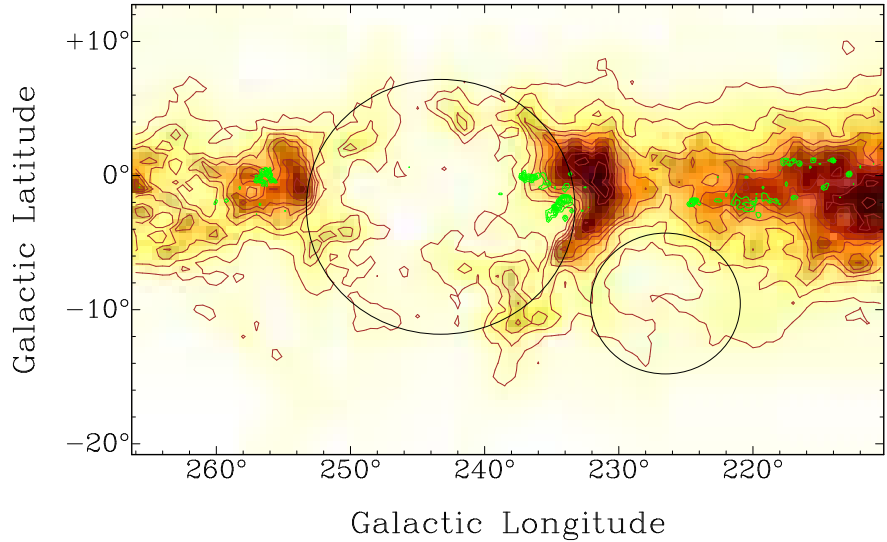}
%
%
\caption{HI emission in the region of GSH243.5-02.5+43.3 (big circle) and GSH226.5-09.5+31.9 (small circle) integrated in the radial velocity interval 30 - 50 km s$^{-1}$ with overlaid CO contours.}
\label{fig3}       
\end{figure}
\end{center}

\begin{center}
\begin{figure}[h]
\includegraphics[scale=.45,angle=0]{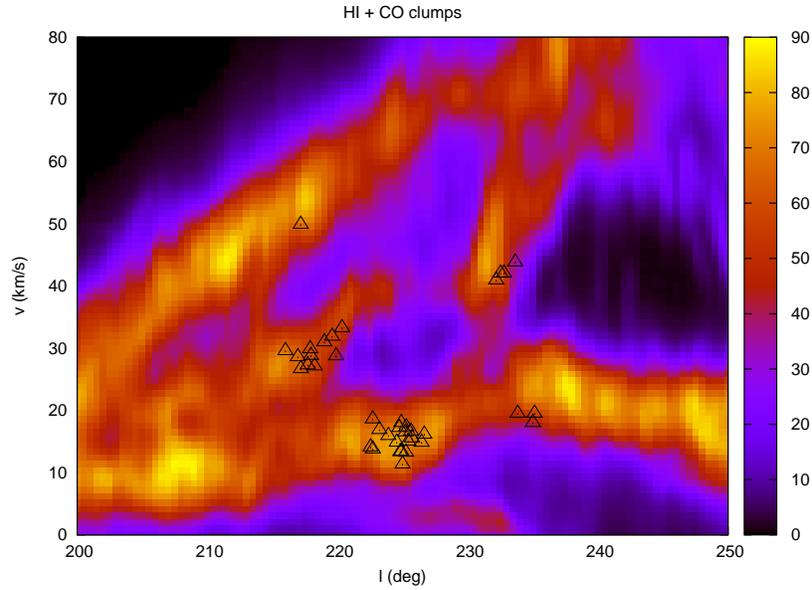}
%
%
\caption{GSH226.5-09.5+31.9: lv diagram. HI emission in the background with triangles showing the CO clumps.}
\label{fig4}       
\end{figure}
\end{center}

\begin{center}
\begin{figure}[h]
\includegraphics[scale=.45,angle=0]{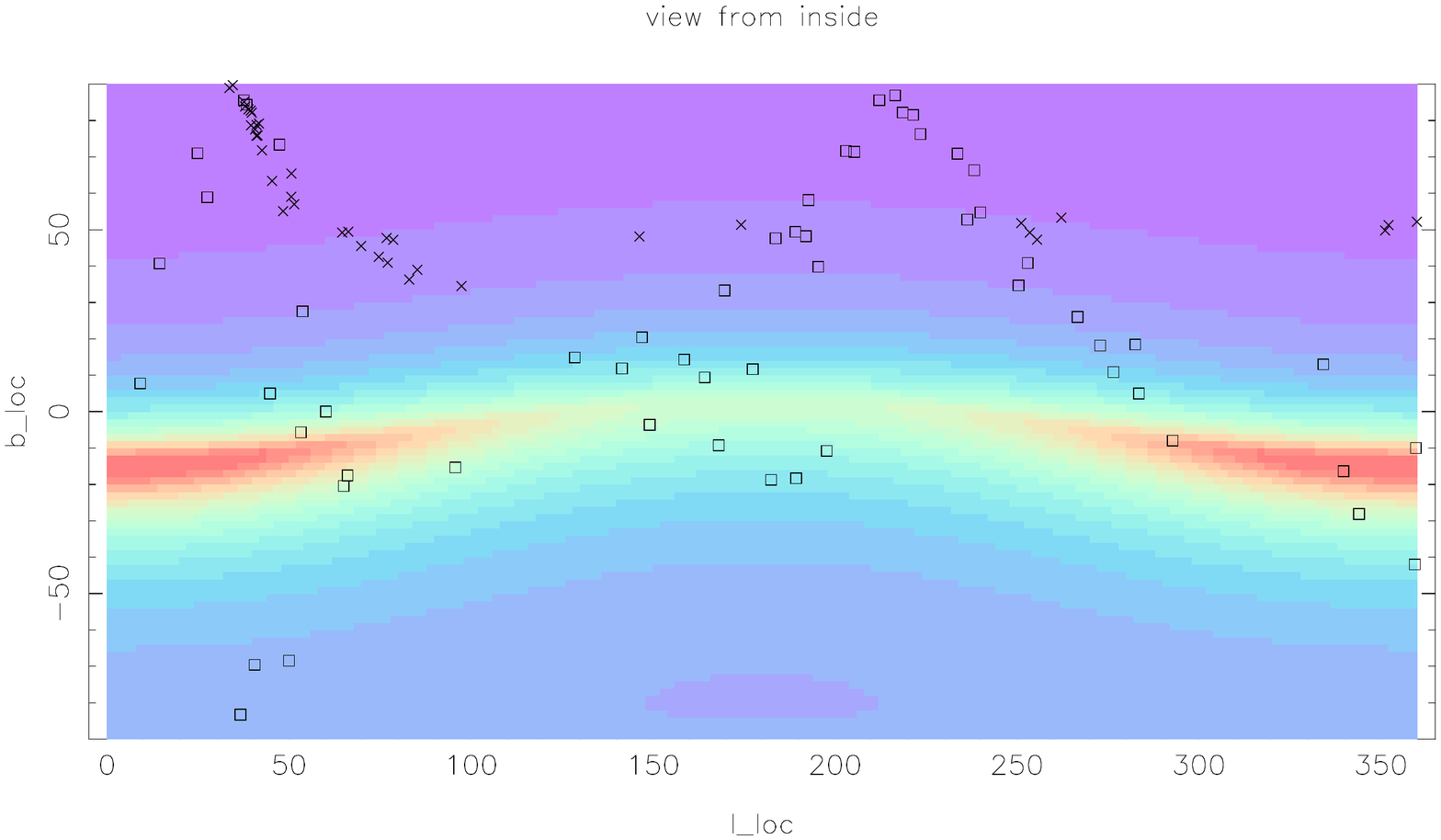}
%
%
\caption{GSH226.5-09.5+31.9: a view from inside. Crosses show the CO and squares HI clumps. }
\label{fig5}       
\end{figure}
\end{center} 

\noindent
Ehlerov\' a and Palou\v s (2005, 2013) analyzed  Leiden/Argentina/Bonn HI survey and discovered a few hundreds of shells (see Fig. \ref{fig2}). Their sizes range up to 1 kpc with a power-law distribution corresponding to the luminosity function of OB associations. Fig. \ref{fig3} shows the (l-b) distribution of HI in the region of the supershell GSH243.5-02.5+43.3 discovered by Heiles (1979) and its neighbouring shell GSH226.5-09.5+31.9 together with overlaid CO emission. In Fig. \ref{fig4} we present the (l-v) diagram of the shell GSH226.5-09.5+031.9. Using the algorithm DENDROFIND (Wunsch et al. 2012), we have identified  HI clumps in  shell walls complemented with CO clumps seen in the CO survey by Dame et al. (2001). In Fig. \ref{fig5} we display the HI and CO clumps in walls of the supershell GSH226.5-09.5+31.9 as they would be seen if the observer resided in its center: they are distributed in a ring inclined relative to the galaxy symmetry plane resembling with its size, the plane-like distribution and of the tilt HI and CO clouds of the Gould's belt. 

The existence of many HI holes in the HI distribution in the Milky Way, some of them expanding with velocities ranging between 10 - 30 km s$^{-1}$, and the correlation of their walls with CO distribution, show that in our Galaxy there may be many Gould's belts. The view from the current position of the Sun, which arrived into the local volume of the local HI hole surrounded by local molecular clouds and  local young OB stars and associations, shows us the structure called Gould's belt. However,  there may be many similar structures along the plane of the Milky Way.  

\section{Origins}

 The formation of molecular clouds  in curling gas flows downstream spiral arms was shown in simulations by Dobbs et al. (2012). There we may see voids surrounded by clouds similar to Gould's belt.    
Star formation occurs in hierarchical patterns (Elmegreen 2011): star forming regions are grouped into complexes of 1 kpc size. This results from turbulent nature of the galaxy ISM, star formation following the dense filaments and knots formed in turbulent cascades. Gould's belt is a star formation complex which is a part of the turbulent cascade downstream from galaxy spiral arms.

\begin{acknowledgement}
This study has been supported by Czech Science Foudation grant 209/12/1795 and by the project RVO: 67985815. This research made use of NASA's Astrophysics Data System. The authors would like to thank Anthony Whitworth for the help with the manuscript.
\end{acknowledgement}
%
%
%


%
%
%

\end{document}